\documentclass[%
 aip,
 amsmath,amssymb,
 reprint,%
]{revtex4-1}

\usepackage{graphicx}
\usepackage{dcolumn}
\usepackage{bm}
\usepackage{mathptmx}

\begin{document}

\preprint{AIP/123-QED}

\title{ Single germanium vacancy centres in nanodiamonds with bulk-like spectral stability}

\author{M. Nahra}
 \affiliation{Laboratory Light, nanomaterials \& nanotechnologies (L2n), CNRS ERL 7004, University of Technology of Troyes, 12 rue Marie Curie, 10004 Troyes Cedex, France}
 \author{D. Alshamaa}
\affiliation{Laboratory of Informatics, Robotics and Microelectronics of Montpellier (LIRMM), UMR 5506, University of Montpellier – CNRS, 34095 Montpellier Cedex, France}
\author{R. Deturche}
 \affiliation{Laboratory Light, nanomaterials \& nanotechnologies (L2n), CNRS ERL 7004, University of Technology of Troyes, 12 rue Marie Curie, 10004 Troyes Cedex, France}
\author{V. Davydov}
 \author{L. Kulikova}
\affiliation{%
 L.F. Vereshchagin Institute for High Pressure Physics, Russian Academy of Sciences, Troitsk, Moscow, 142190, Russia}%
\author{V. Agafonov}
\affiliation{GREMAN, UMR CNRS CEA 7347, Université de Tours, 37200 Tours, France}%
\author{C. Couteau}
\email{christophe.couteau@utt.fr}
  \affiliation{Laboratory Light, nanomaterials \& nanotechnologies (L2n), CNRS ERL 7004, University of Technology of Troyes, 12 rue Marie Curie, 10004 Troyes Cedex, France}
\date{\today}

\begin{abstract}
Motivated by the success of group IV colour centres in nanodiamonds (NDs) for hybrid technology requiring a single photon source, we study single germanium-vacancy (GeV$^-$) centres in NDs at room temperature with size ranging from 10 to 50 nm and with remarkable spectral properties. We characterize their zero-phonon line (ZPL), study their internal population dynamics and compare their emission properties in the framework of a three level model with intensity dependent de-shelving. Furthermore, we characterize their lifetime, polarization and brightness. We find a maximum photon emission count rate of 1.6 MHz at saturation. We also report a polarization visibility of 92\% from the fluorescence light, which potentially makes GeV$^-$ centres good candidates for quantum key distribution (QKD) requiring polarized single photons. We show that the GeV$^-$ in NDs presented in this work have a comparable spectral stability compared to their bulk counterpart which is needed for future applications using nanodiamonds.


\end{abstract}

\maketitle

\section{Introduction}
Single defect centres in diamond have become a highly attractive candidate for solid-state single photon source applications  \cite{aharonovich2011diamond,ladd2010quantum,childress2014atom}. Nowadays, many studies have been carried out on nitrogen vacancy (NV$^-$) and silicon vacancy (SiV$^-$) colour centres showing some limitations. Despite the remarkable electronic spin coherence of NV$^-$ centres\cite{balasubramanian2009ultralong}, they suffer from a low optical emission of only 4\% into the ZPL, as well as a polar symmetry rendering it sensitive to external perturbations\cite{doherty2013nitrogen}. In contrast, SiV$^-$ centres offer exceptional optical properties with a narrow linewidth, 80\% of the emission concentrated in the ZPL and a small phonon side band\cite{wang2005single}. More importantly, the SiV$^-$ inversion symmetry makes it robust against external perturbations such as, crystal strain and fluctuating magnetic and electric field \cite{hepp2014electronic}. However, it suffers from low quantum efficiency and low coherence time due to non-radiative processes limited by phonon-mediated transitions \cite{neu2011single,jahnke2015electron}. In addition, colour centres in bulk diamond suffer from low collection efficiency due to the high refractive index of diamond\cite{rogers2014multiple,babinec2010diamond,englund2010deterministic}. Diamond nanostructures can overcome this issue. In particular, small NDs have reduced light scattering and they can be more easily coupled to photonic structures which makes them more appropriate for hybrid technologies \cite{siampour2018chip,fehler2020purcell}. However, colour centres in NDs present a large spectral distribution of their ZPL, which is a major drawback for such emitters. Strain and charge fluctuations at the surface of small NDs are responsible for this random spectral distribution  \cite{neu2011single,neu2013low,choi2018varying, lindner2018strongly}.

 To overcome these issues, research has been oriented towards a more efficient and bright colour centre in small NDs, the GeV$^-$ centre\cite{bhaskar2017quantum}. This defect is a promising colour centre with superior optical properties. It possesses the same inversion symmetry than the SiV$^-$ colour centre\cite{goss2005vacancy,iwasaki2015germanium} thus having equivalent optical properties\cite{palyanov2015germanium}. However, it outperforms the SiV$^-$ colour centre with its higher quantum efficiency \cite{bhaskar2017quantum}. 


In this article, we study single GeV$^-$ centres in NDs with a mean size of about 20 nm, grown using the high-pressure-high-temperature (HPHT) method. First, we analyze their ZPL properties, showing a stable ZPL with a small spectral shift as compared to their bulk counterpart. We then demonstrate the single photon nature of the emitted light from single NDs using an Hanbury Brown and Twiss setup. Then, in order to better understand the emission properties, we proceed by studying the internal population dynamics of two typical single GeV$^-$ centres from two distinct NDs. Our study is analysed within the framework of an extended three level system model with intensity dependent de-shelving\cite{neu2012photophysics}. Finally, we characterize their full optical properties such as brightness, lifetime and polarization. 


\begin{figure*}
\centering
\includegraphics[trim=0cm 0cm 0cm 0cm,width=0.85\linewidth,keepaspectratio]{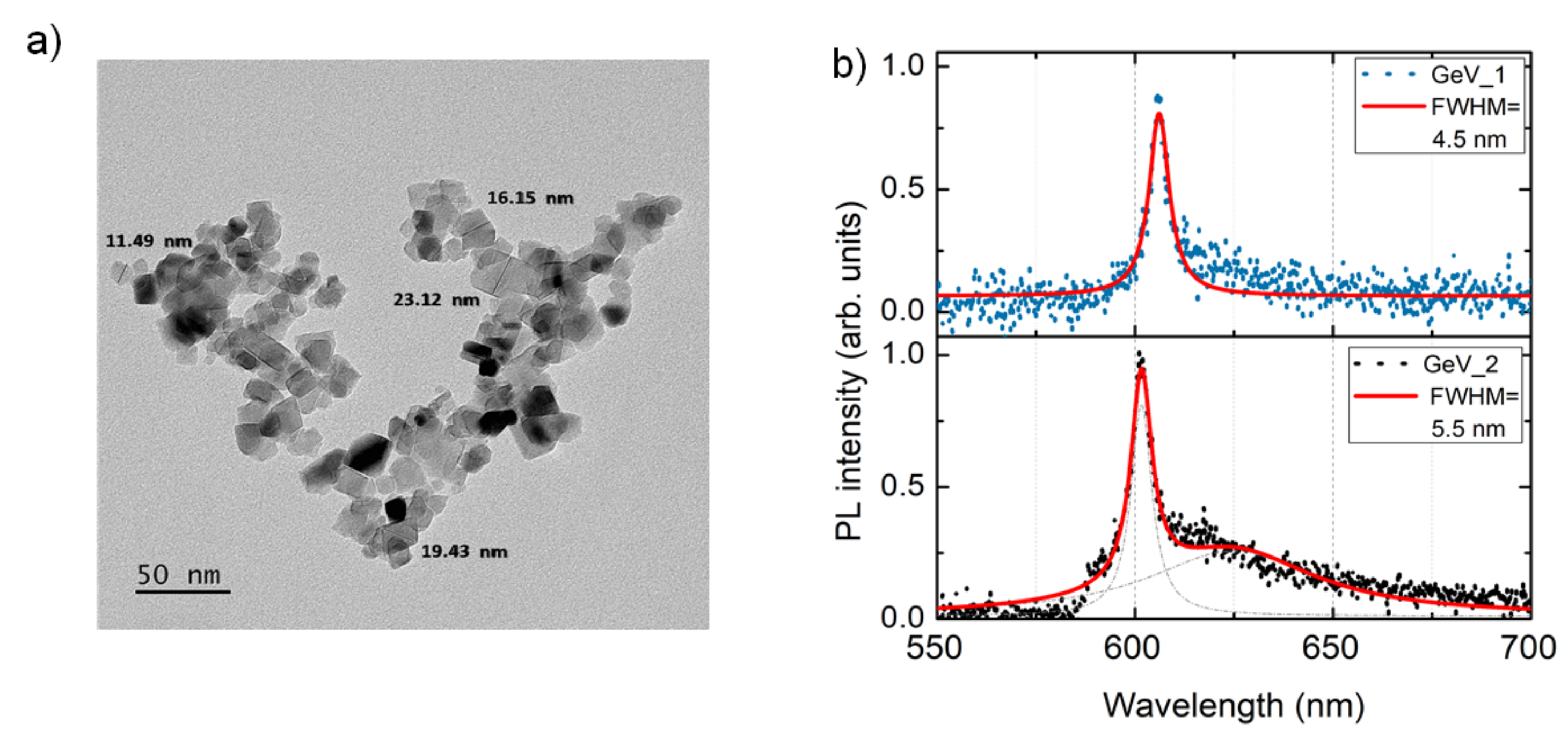}
\qquad \qquad
\\ 
\includegraphics[trim=0cm 0cm 0cm 0cm,width=0.75\linewidth,keepaspectratio]{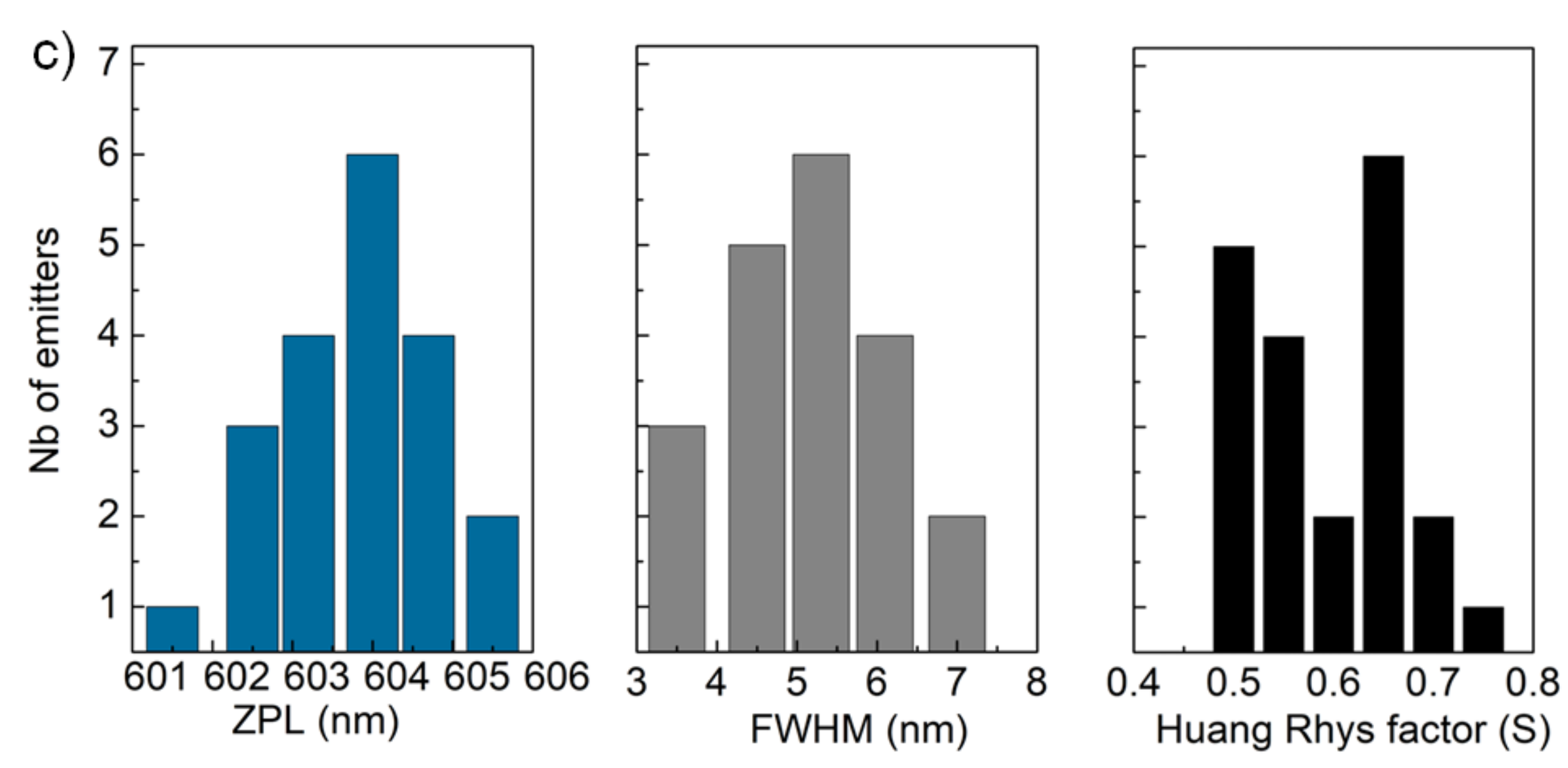}
\caption{\label{optical} \textbf{Characterization of single GeV centres}. \textbf{\ref{optical}a} TEM image for HPHT nanodiamonds taken after ultrasonic treatment. Their size average is 20 nm.\textbf{\ref{optical}b}
Photoluminescence (PL) spectrum for two GeVs centres. The red line denotes a Lorentzian fit of the ZPL for GeV1 while it denotes a multi-peak Lorentzian fit of both the ZPL and PSB for GeV2 . \textbf{\ref{optical}c} Histogram distributions of ZPL, FWHM and Huang Rhys factor for 20 emitters.} 
\end{figure*}

\section{METHODS}
We studied several GeV$^-$ centres in diamond nanocrystals produced using the HPHT technique. 
To form the complex GeV$^-$ centre, tetraphenylgermanium (C$_{24}$H$_{20}$Ge) was introduced during the growth process and incorporated into the diamond nanocrystals. After that, and in order to remove the excess of Ge and germanium oxide (GeO$_2$), the NDs were suspended in hydrofluoric acid (HF) at 160$^\circ$C for 2 hours. Later on, they were dissolved in a  solution of ultrapure water and isopropanol. Next, they were ultrasonicated to disperse the nanodiamonds and finally spin coated on a silicon substrate. The detailed growth technique can be found in the Supplementary Materials S1 . Figure \ref{optical}a shows a transmission electron microscopy (TEM) image of some GeV$^-$ centres in NDs  of different sizes ranging from 10 to 50 nm, with an average of 20 nm taken after the chemical and ultrasonic treatment.
In order to investigate the optical properties of the GeV$^-$ centres in NDs and identify whether a centre is single or not, the sample was characterized by optical micro-photoluminescence (PL) at room temperature. The experimental setup used for the optical characterization is shown in Supplementary Materials S2. The GeV$^-$ centres were excited using an off-resonantly continuous wave laser at 532 nm. The laser light was focused through a 100x dry microscope objective and the fluorescence was collected using the same objective. The collected fluorescence light was filtered by a narrow band-pass with a full-width at half maximum (FWHM) of 14 nm around 600 nm in order to avoid Raman signal of diamond and other background emission from the substrate and the carbon matrix of the NDs. After that, it was directed to one of two paths: a spectrometer for photoluminescence measurements or a Hanbury Brown and Twiss  set-up for second order correlation function (g$^{(2)}(\tau)$). For the purpose of polarization measurement, a linear polarizer was added to the detection path and rotated from 0 to 360 $^\circ$.

\begin{figure*}
  \centering
         \includegraphics[trim=0cm 0cm 0cm 0cm, clip=true,width=0.42\textwidth]{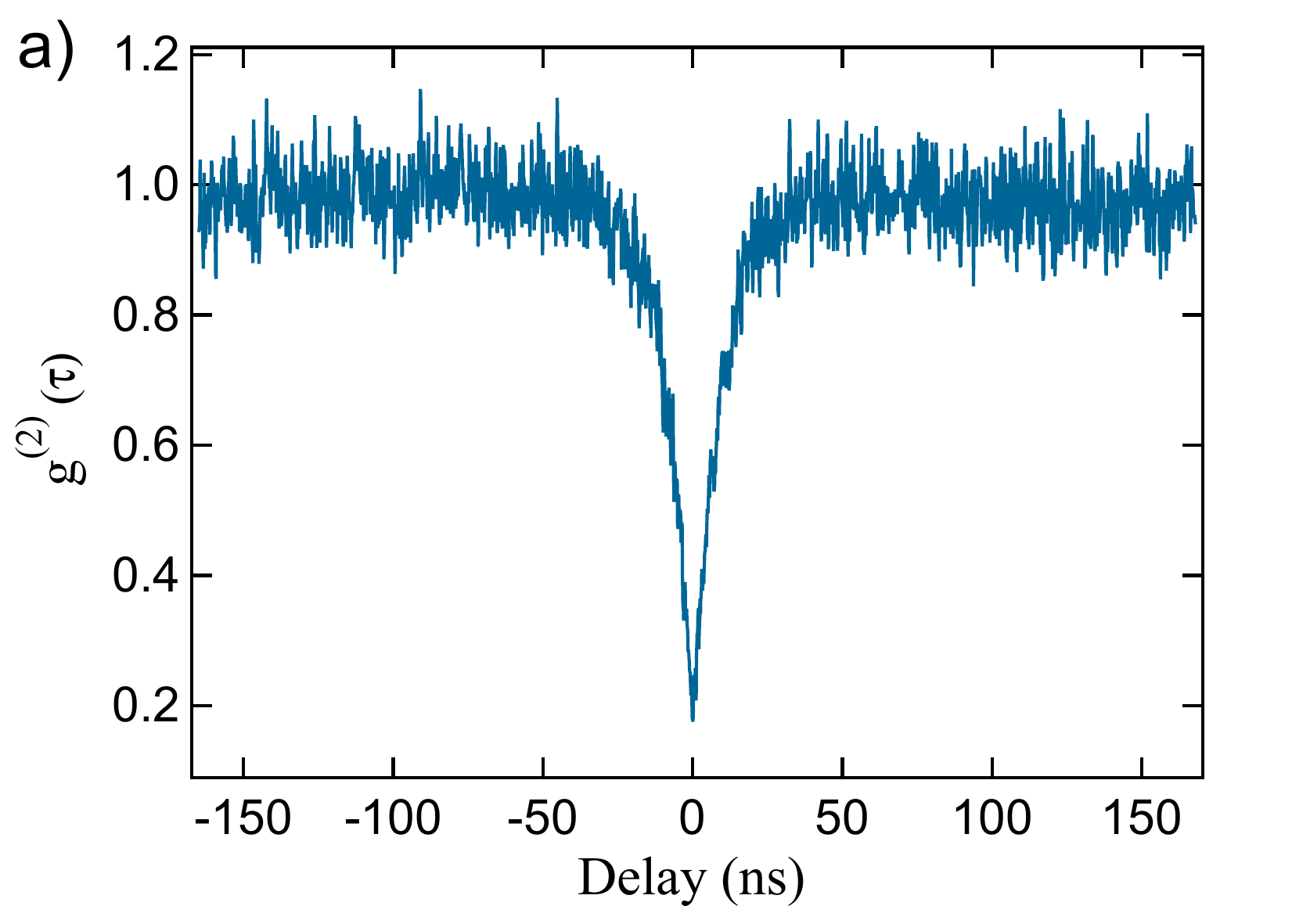}\label{continuous}
        \qquad
        \includegraphics[trim=0cm 0cm 0cm 0cm, clip=true,width=0.42\textwidth]{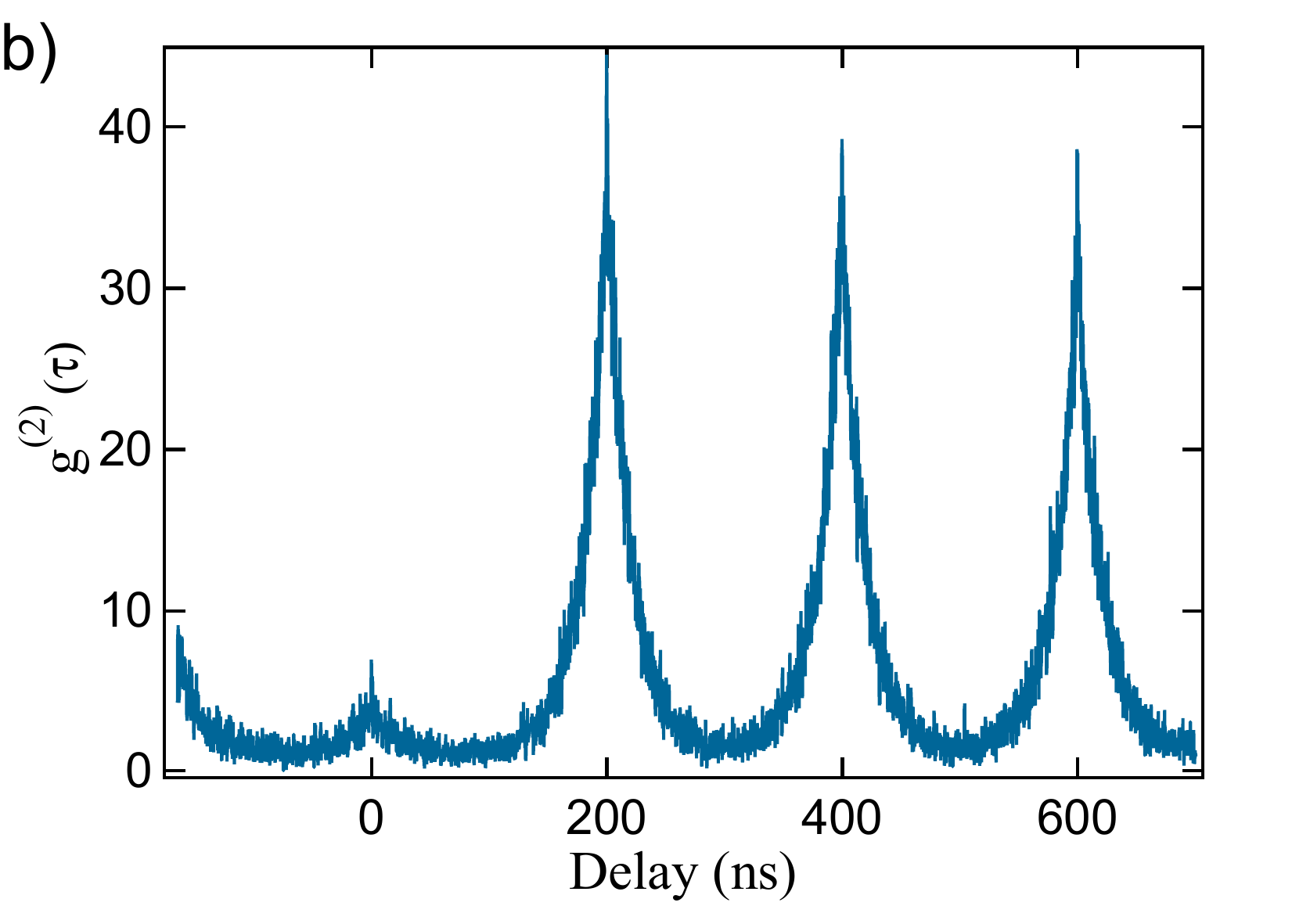}\label{pulsed}
   \caption{\label{g2pulse}\textbf{Second order correlation function for GeV2}. \textbf{\ref{g2pulse}a} with a continuous excitation laser and \textbf{\ref{g2pulse}b} with a pulsed excitation laser. A value of of less than 0.5 is a clear evidence for the single photon nature of the emitted luminescence.}
    \end{figure*}

\section{Results and discussions}
\subsection {Photoluminescence spectra}

In order to study the optical properties of our GeV$^-$ colour centres, we first performed PL measurements at room temperature. Figure \ref{optical}b shows normalised PL spectra for two single GeV$^-$ centres with a ZPL centred at 605.5 and 601.5 nm respectively. Using a Lorentzian fit of the ZPL, we measure a FWHM of 4.5 for GeV1. However for GeV2, we use a multi-peak Lorentzian fit and measure a FWHM of 5.5 nm for the ZPL. This is comparable with the observations of a FWHM of 5.5 nm in bulk diamond at room temperature \cite{palyanov2015germanium}. 
We note that the FWHM around 5 nm is supposed to be caused by the isotopic shift of the ZPL due to the presence of Ge isotopes  \cite{palyanov2015germanium,ekimov2017anharmonicity}. 
Moreover, we measure the Huang Rhys factor $S$ defined by $\frac{I_{ZPL}}{I_{tot}}=e^{-S}$ in order to determine the strength of electron-phonon coupling \cite{walker1979optical}. A value of the $S$ factor close to zero corresponds to a pure electronic emission without phonon coupling, while a value higher than one implies a strong electron-phonon coupling. This is typically the case for NV$^-$ centres with $S$=3.73\cite{davies1981jahn}. In our case, we obtained $S$ factors of 0.5 and 0.79 for GeV1 and GeV2 respectively. These values are lower than other values reported for GeV of 1.3 \cite{haussler2017photoluminescence}. 
These measurements were carried out on 20 centres. It was found that these centres exhibit almost the same ZPL, with a slight variation of $\pm$2 nm around 603.5 nm. It was also found that the FWHM varies from 4 and 7 nm. In addition, an average $S$ factor of $S_{mean}$ = 0.65 was reported. 
Figure \ref{optical}c presents the distribution of the ZPL, FWHM and $S$ factor for 20 studied centres which allowed us to come up with mean values of 603.5 nm, 5.2 nm and 0.65 respectively.
Due to the absence of studies on GeV$^-$ centres in NDs, we compare the stability of ZPL to that of SiV$^-$ in NDs. Previous studies reported a wide spread of the position of the ZPL of SiV in NDs \cite{neu2011single,lindner2018strongly}. This is attributed to the strain effect within the NDs. Other studies demonstrated that a special surface treatment of NDs with hydrogen-plasma or oxygen improves the spectral stability of the SiV$^-$ ZPL \cite{rogers2019single,neu2013low}. It is worth mentioning that no particular surface treatment was performed on our NDs.
The unusually narrow distribution of the ZPL position in the PL spectra of GeV$^-$ centres in the studied NDs indicates low spectral diffusion. This spectral diffusion is caused, in our view, by the low internal strain of NDs obtained on the basis of hydrocarbon growth systems in the (P,T) region of the thermodynamic stability of diamond at a pressure of 8.0 GPa and at rather high temperatures of $\approx$ 1500$^\circ$C, which facilitates the efficient annealing of structural defects in diamond. An additional factor that reduces the number of electroactive surface defects is probably the fluorination of the surface of nanodiamonds during their preliminary chemical treatment according to the following chemical reaction Diam-OH + HF $\longrightarrow$ Diam-F + H$_2$O. 
These properties render our GeV$^-$ centres potentially suitable for quantum technologies as they have the potential to provide indistinguishable single photons.


\subsection{Second order correlation function g$^{(2)}$} 
\begin{figure*}

\includegraphics[trim=0cm 0cm 0cm 0cm, clip=true,width=0.42\linewidth,keepaspectratio]{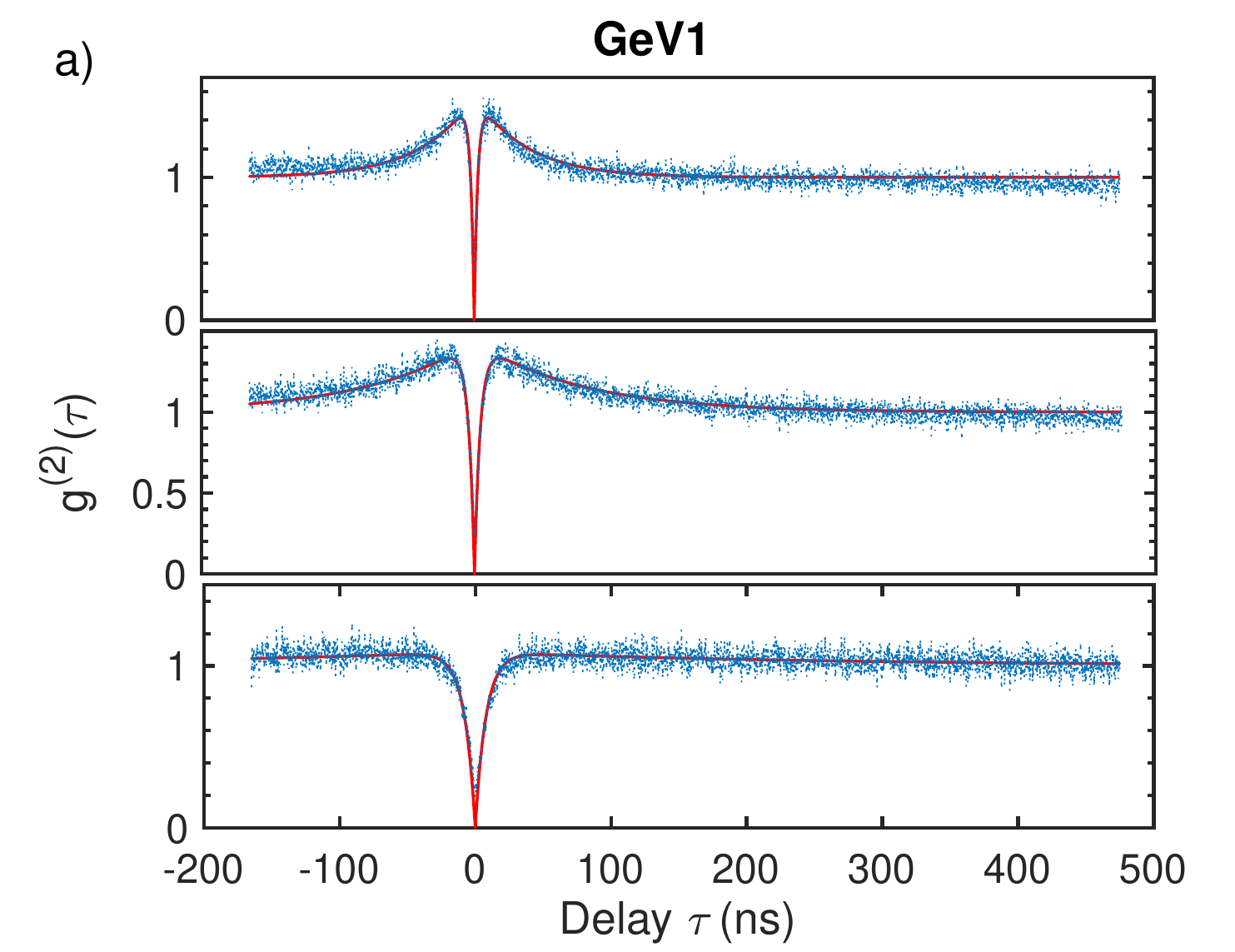}\label{g2-12}
\qquad
\includegraphics[trim=0cm 0cm 0cm 0cm, clip=true, width=0.42\linewidth,keepaspectratio]{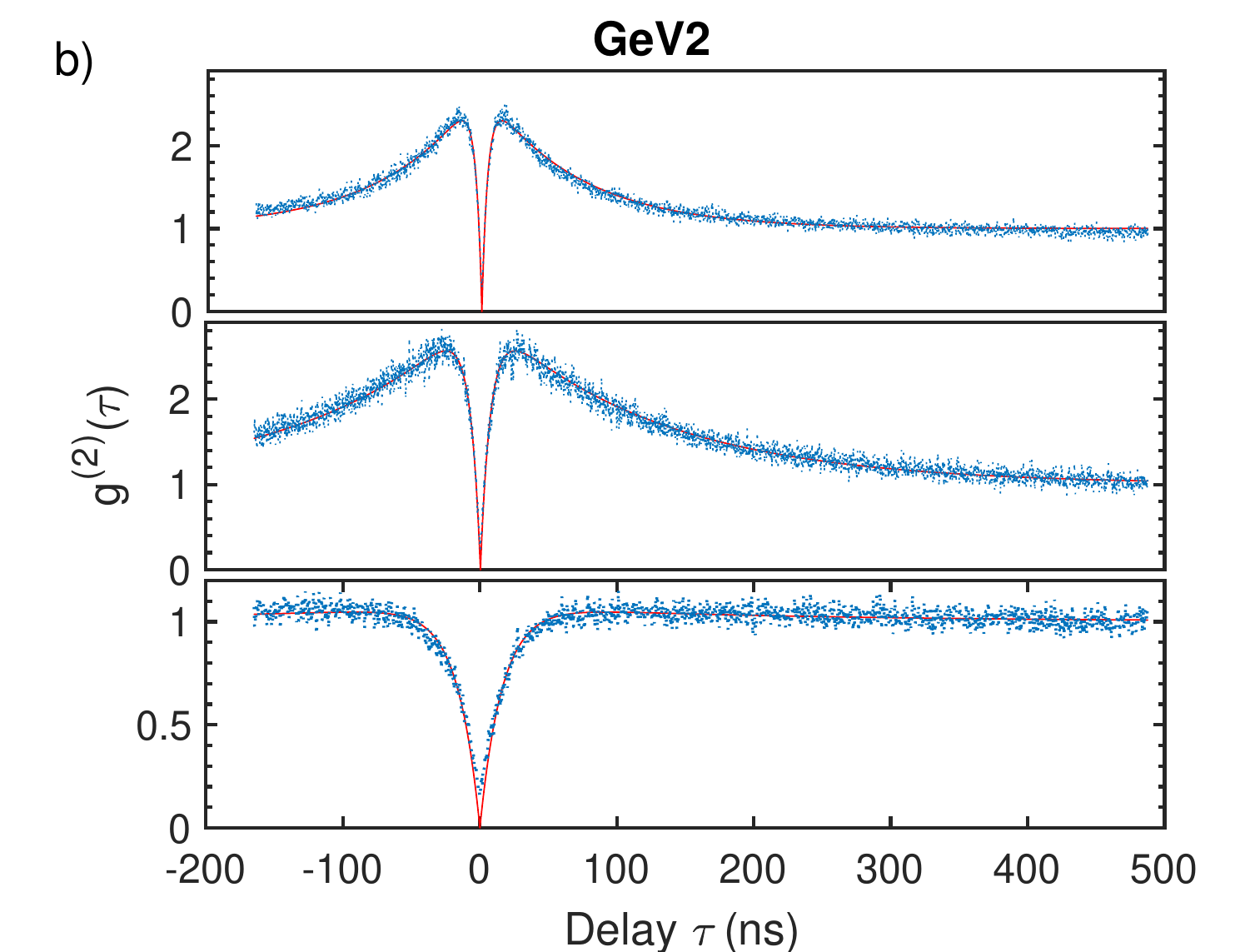}\label{g2-105}
\\
\includegraphics[trim=0cm 0cm 0cm 0cm, clip=true,width=0.42\linewidth,keepaspectratio]{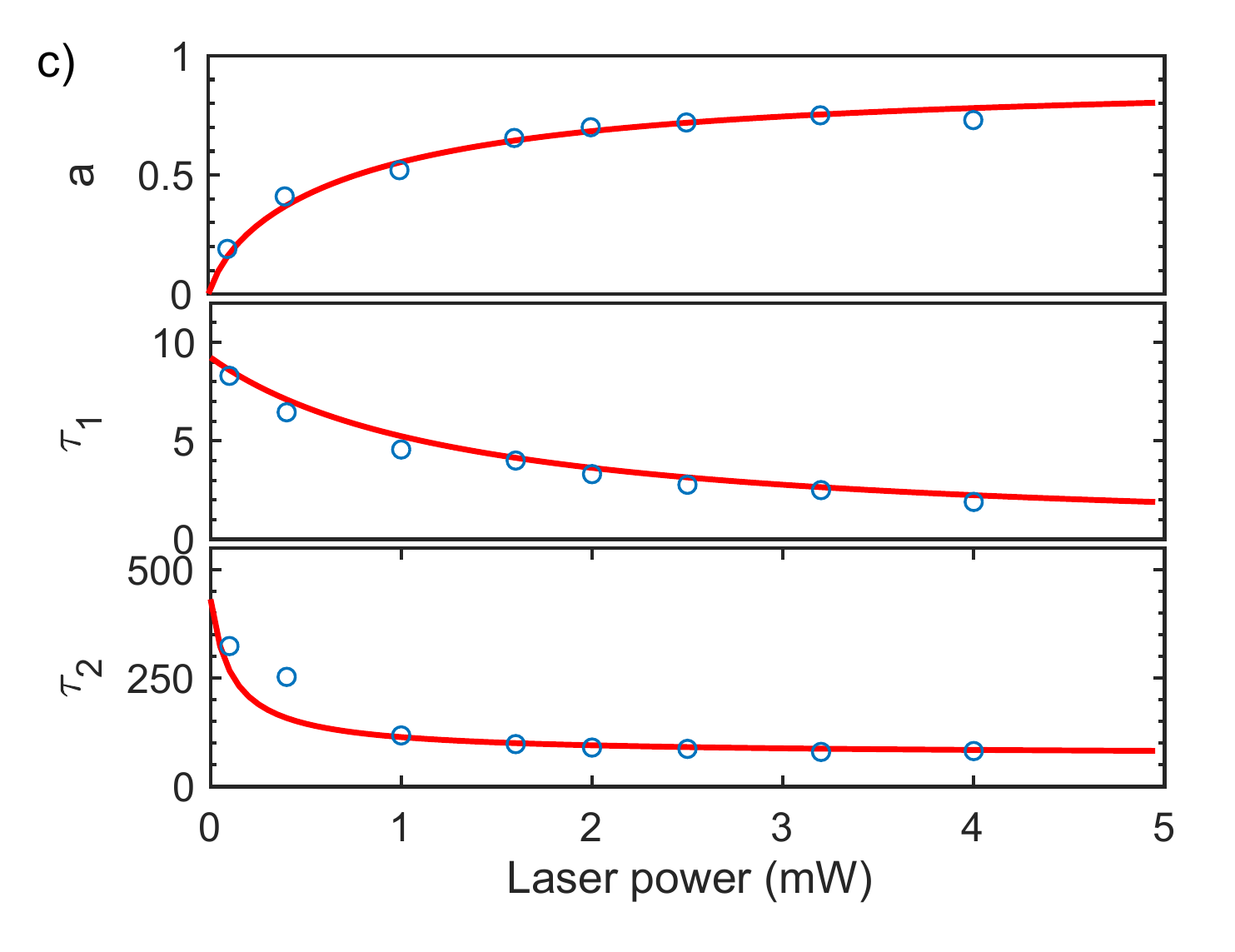}\label{a,tau-12}
 \qquad 
\includegraphics[trim=0cm 0cm 0cm 0cm, clip=true, width=0.42\linewidth,keepaspectratio]{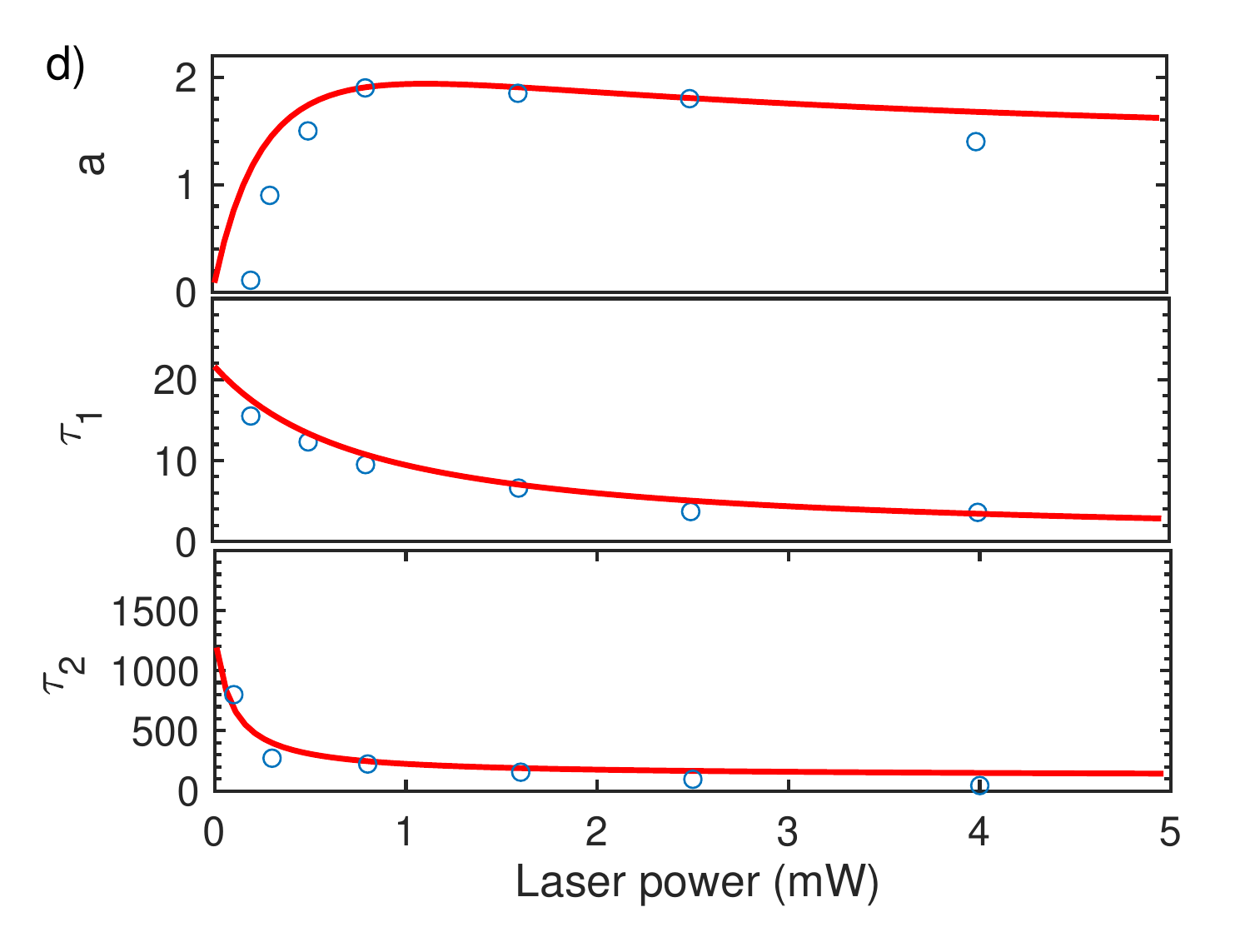}\label{a,tau-105}
\caption{\label{model} \textbf{g$^{(2)}$ modelling according to a three level system with intensity dependent de-shelving}. \textbf{\ref{model}a, \ref{model}b} Three representative normalised autocorrelation curves taken at different excitation laser powers 0.1, 1 and 4 mW for GeV1 and GeV2 respectively. \textbf{\ref{model}c, \ref{model}d} Power dependence of the fitting parameters a, $\tau_1$ and $\tau_2$ for GeV1 and GeV2 respectively.}
\end{figure*}

In order to investigate the presence of single photons emitted by a single GeV$^-$ centre as well as to understand its internal population dynamics, we measured the intensity autocorrelation function $g^{(2)}$ of some of our single defect centres. An example of a $g^{(2)}$ result under a pulsed and continuous laser excitation is shown in figure \ref{g2pulse}. The obtained value of $g^{(2)}=0.2$ at $\tau$ = 0 ns is a clear evidence of a single photon source from the GeV$^-$ defect emission. In the following, we aim to study the internal population dynamics of our GeV$^-$ centres using a continuous laser at various excitation powers. To this end, we considered seven GeV$^-$ centres in distinct NDs and measured their intensity correlation function $g^{(2)}$. Six out of the seven studied GeV centres have presented almost identical properties. We made the choice to show in this article the results of one of these six typical GeV$^-$ centres, namely GeV1. We also present results from the seventh GeV$^-$ centre, namely GeV2, that we studied more thoroughly as its properties were different from the other 6 ones as we will show.  Figures \ref{model}a and \ref{model}b present normalised $g^{(2)}$ functions measured at different excitation powers. The $g^{(2)}$ does not follow a Poissonian source over all the delay time. Rather, at intermediate delay, the $g^{(2)}$ increases and then decreases to become approximately flat for a very long delay time. Hence, in order to normalize the $g^{(2)}$ functions, we set their long delay limit to 1. The pronounced photon antibunching at zero time delay clearly proves the single photon nature of the light emitted by these two GeVs centres under investigation. We point out here that the data of $g^{(2)}$ are given without any correction factor from background emission . As one could expect, for longer time delays, a photon bunching can be noticed where the value of $g^{(2)}$ exceeds one. The presence of bunching and antibunching allows us to represent the GeV$^-$ colour centre emission by a three level model comprising a ground state (1), an excited state (2) and a metastable state (3). 
The $g^{(2)}$ function for an ideal three level system can be expressed as follows: 
\begin{equation} \label{eqantibunching}
    g^{(2)}(\tau)=1-(1+a)e^{-\frac{ \mid \tau  \mid}{\tau_{1}}}+ae^{-\frac{\mid  \tau \mid}{\tau_{2}}} ,
\end{equation}
where $\tau_{1}$ and $\tau_{2}$ designate the antibunching and bunching time constants respectively and with $a$ the so-called degree of bunching. The details of the employed $g^{(2)}$ function can be found here \cite{kurtsiefer2000stable}. 
This typical function reflects the population dynamics of state 2. The parameters $a$, $\tau_{1}$ and $\tau_{2}$ can be obtained via solving the rate equations for the populations in the system \cite{wang2007solid}, as shown in the following set of equations:

\begin{align}
    \tau_{1,2} &=\frac{2}{A\pm\sqrt{A^{2}-4B}} \label{eq2}\\ 
    A &=k_{12}+k_{21}+k_{23}+k_{31} \label{eq3}\\
    B &= k_{12}k_{23}+k_{12}k_{31}+k_{21}k_{31}+k_{23}k_{31} \label{eq4}\\
    a &=\frac{1-\tau_{2}k_{31}}{k_{31}(\tau_{2}-\tau_{1})\label{eq5}}
\end{align}

where $k_{ij}, i,j\in \{1,2,3\}$ represent the rate coefficients from state $i$ to state $j$. After replacing A and B in equation (\ref{eq2}) by their expressions in equations (\ref{eq3}) and (\ref{eq4}) and simplifying, the time constant $\tau_2$ can be expressed as $\tau_2 = \frac{1}{k_{31}}$. In this model, all the rate coefficients are intensity-independent except the pumping rate $k_{12}=KP$ which depends linearly on the excitation power. However, when we consider $k_{31}$ to be constant, the model fails to fit our experimental data. As figures  and  show, $\tau_2$ is intensity dependent, which means that $k_{31}$ should not be constant. 

\begin{figure}
   \centering
    \includegraphics[width=0.9\linewidth] {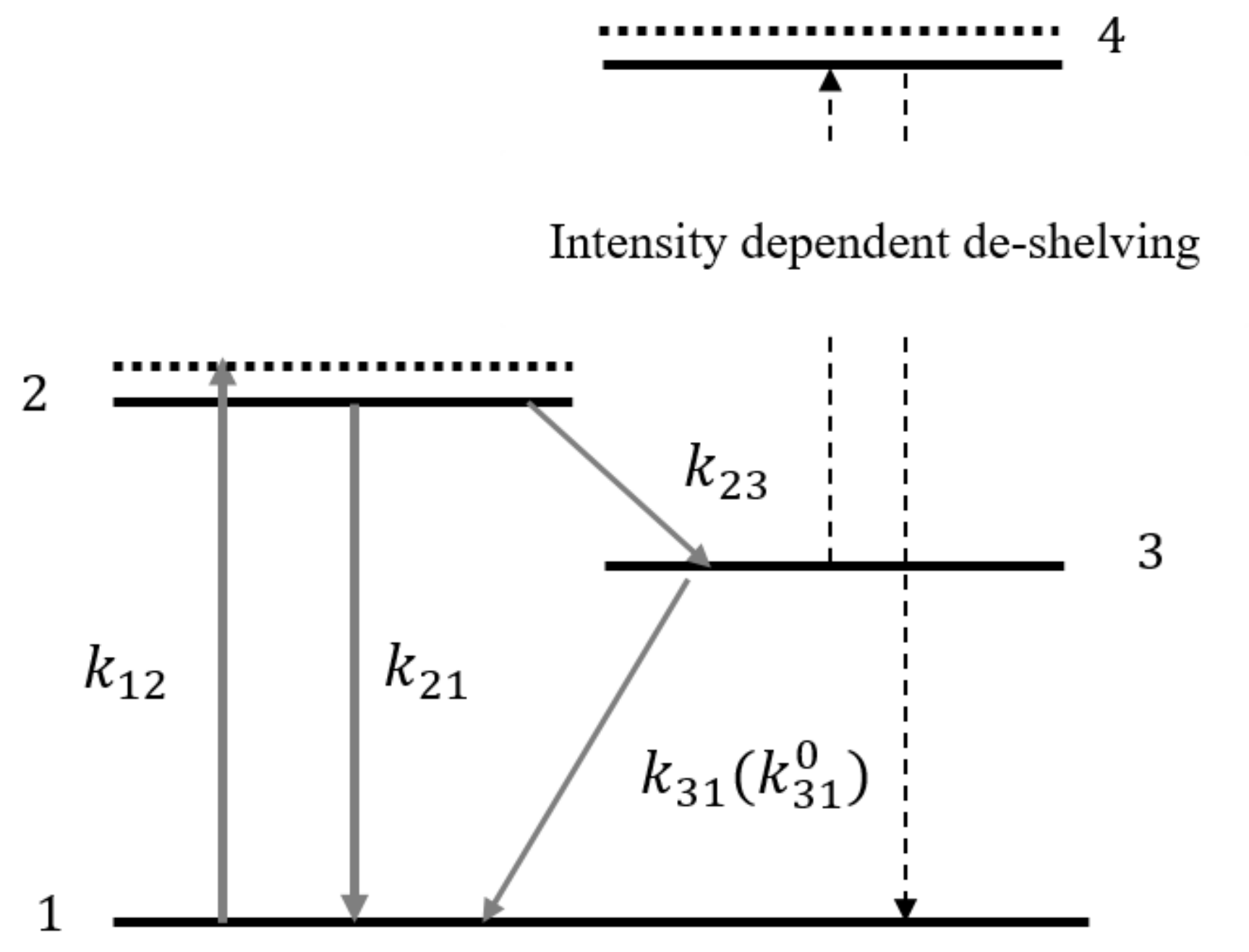}
    \caption{Three level system with intensity dependent de-shelving used to study the population dynamics of GeV$^-$ colour centres in this article.}
    \label{4level}
\end{figure}

For that reason, we extend our model to a 3 level system comprising an intensity dependent de-shelving state. Figure \ref{4level} illustrates the three level system model with intensity dependent de-shelving. Here, $k_{31}$ follows a saturation law as shown by the following equation:
\begin{equation}
    k_{31} = \frac{A_1P}{P+ B_1} + C_1,
\end{equation}

 where $A_1= \frac{1}{\tau_2^\infty}$ and $C_1= \frac{1}{\tau_2^0}$ represent the high and low power limits of the de-shelving process respectively and $B_1$ represents the saturation power of the de-shelving state. A similar model with intensity dependent de-shelving was used in the past for single molecules \cite{treussart2001photon}, NE8 centres \cite{wu2006narrow} and SiV colour centres \cite{neu2011single}.  
In order to determine the model parameters  $A_1$, $B_1$, $C_1$, $K$ and $k_{21}$ and $k_{23}$, an optimisation algorithm was developed to fit the model to the experimental data at a given excitation power. The same obtained parameters were employed in our model to fit the experimental data at all excitation powers. Table \ref{table1} summarizes the obtained fitting parameters for the two studied GeVs. Details of the optimisation algorithm and the calculation of the parameters can be found in Supplementary Materials S3. 
\begin{table}[h!]
\caption{\label{table1} Optimised fitting parameters deduced using the extended three level model. All parameters $A_1$, $C_1$, $k_{21}$ and $k_{23}$ are expressed in GHz while $B_1$ is expressed in mW.}
\begin{ruledtabular}
\begin{tabular}{lccccr}
GeV$^-$ & $A_1$ (GHz)&$B_1$ (mW)& $C_1$ (GHz) &$k_{21}$ (GHz) &$k_{23}$  (GHz)\\
\hline
1 & 0.0051 & 0.45 & 0.0022 & 0.1014& 0.0065  \\
2 & 0.002 & 1.42 & 0.0007 & 0.0458 & 0.0052 \\
\end{tabular}
\end{ruledtabular}
\end{table}
Figures \ref{model}c and \ref{model}d show the experimental and fitting data of the variation of the parameters $a$, $\tau_{1}$ and $\tau_{2}$ as a function of the excitation power. The value of $a$ for GeV2 reaches 2 and is higher than that for GeV1 that is less than 1. This means that GeV2 has a stronger coupling to the metastable state than GeV1. As mentioned earlier, this behaviour for GeV2 was found to be different than the other 6 GeV$^-$ centres studied (including GeV1). 
 The value of the ratio $\frac{k_{23}}{k_{31}}$ gives us the probability of transition to the metastable state. which is lower for GeV1 
 We noticed a ratio of 1.1 for GeV1 lower than that of GeV2 equal to 2.6. This result is in agreement with what we obtained for the bunching parameters $a$. The difference in the coupling to the metastable state, which in turn affects the brightness of the GeV$^-$, might be due to a different concentration of nitrogen donor in each ND \cite{thiering2018ab}. Our choice of these 2 GeVs is based on the significant difference in terms of their population dynamics. These, in turn, affect the optical properties of the GeV$^-$ centres such as saturation and lifetime, as will be discussed in details in the following sections.  


\begin{figure*}
\includegraphics[trim=0cm 0cm 0cm 0cm,width=0.35\linewidth,keepaspectratio]{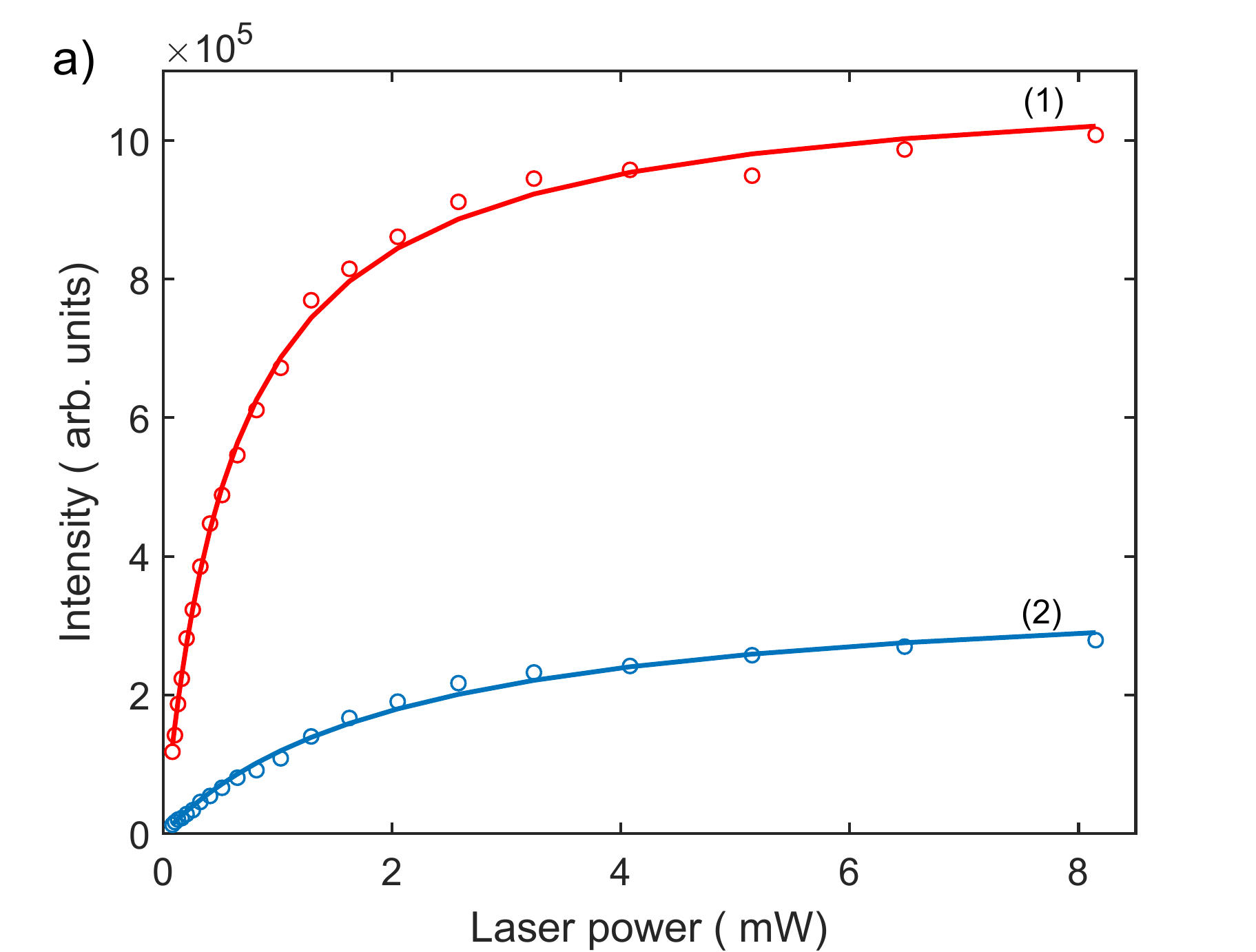}\label{saturation}
\qquad
\includegraphics[trim=0cm 0cm 0cm 0cm, clip=true, width=0.35\linewidth,keepaspectratio]{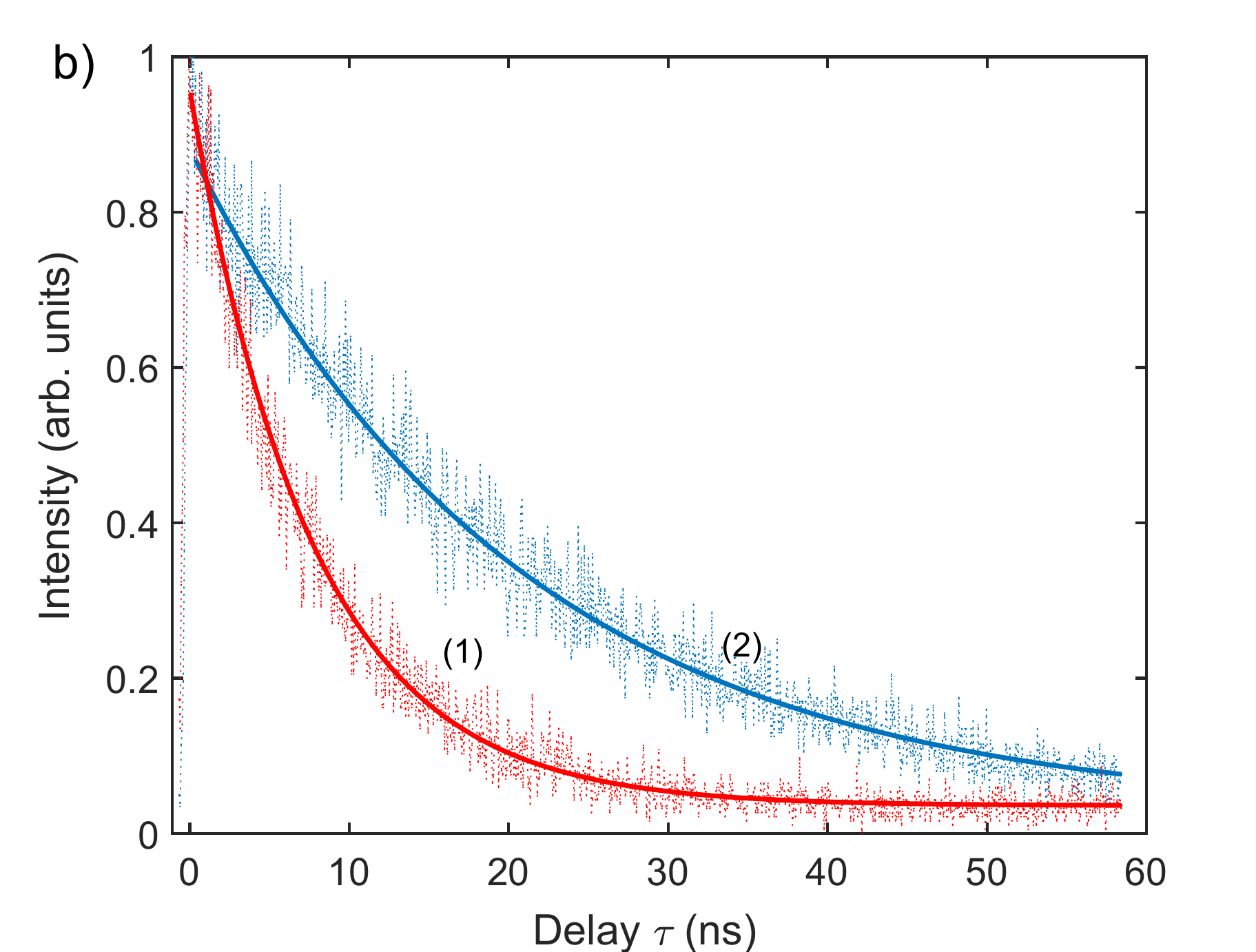}\label{lifetime}
\\
\includegraphics[trim=0cm 0cm 0cm 0cm, clip=true, width=0.32\linewidth,keepaspectratio]{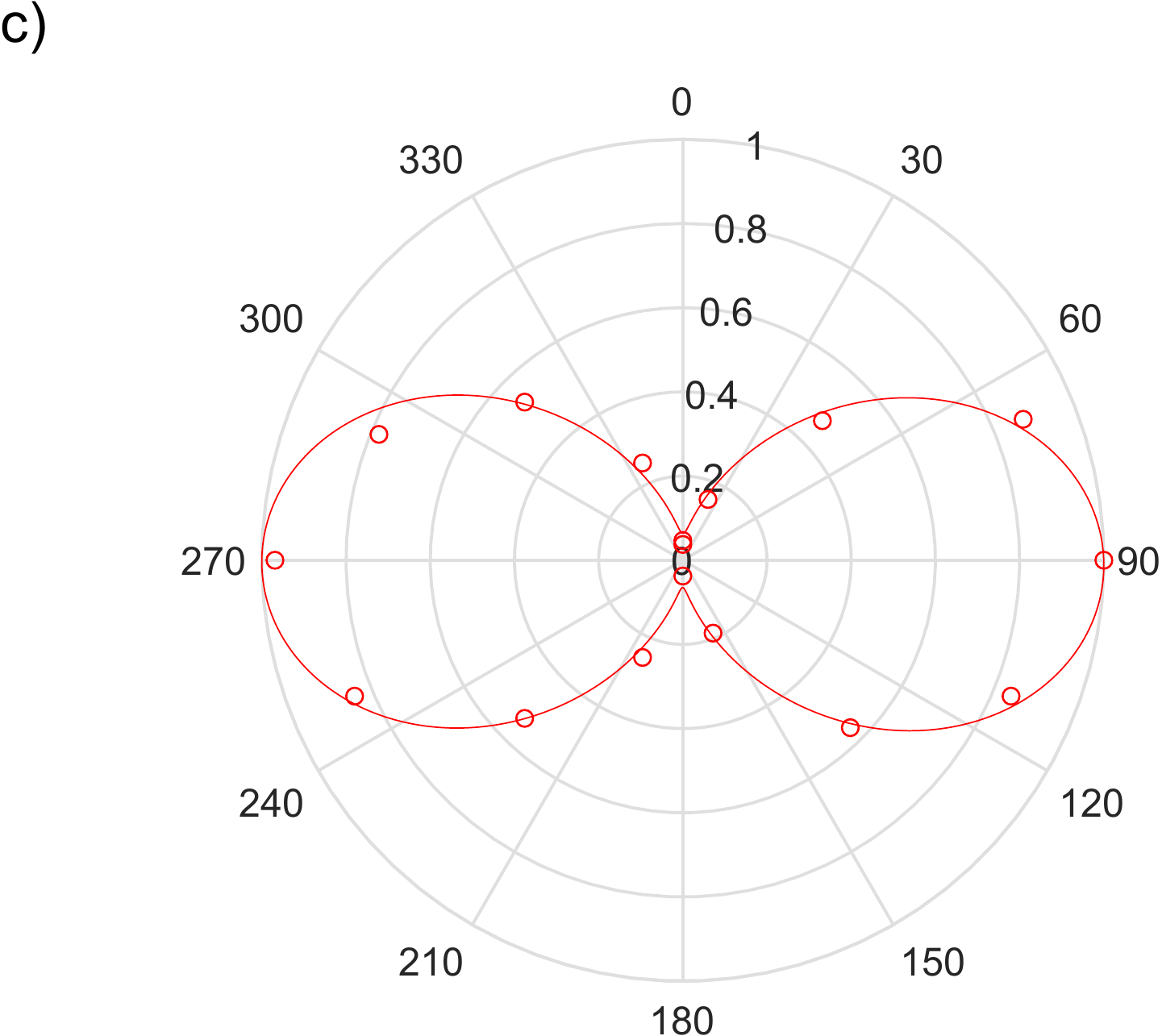}\label{sin1}
\qquad\qquad
\includegraphics[trim=0cm 0cm 0cm 0cm, clip=true, width=0.32\linewidth,keepaspectratio]{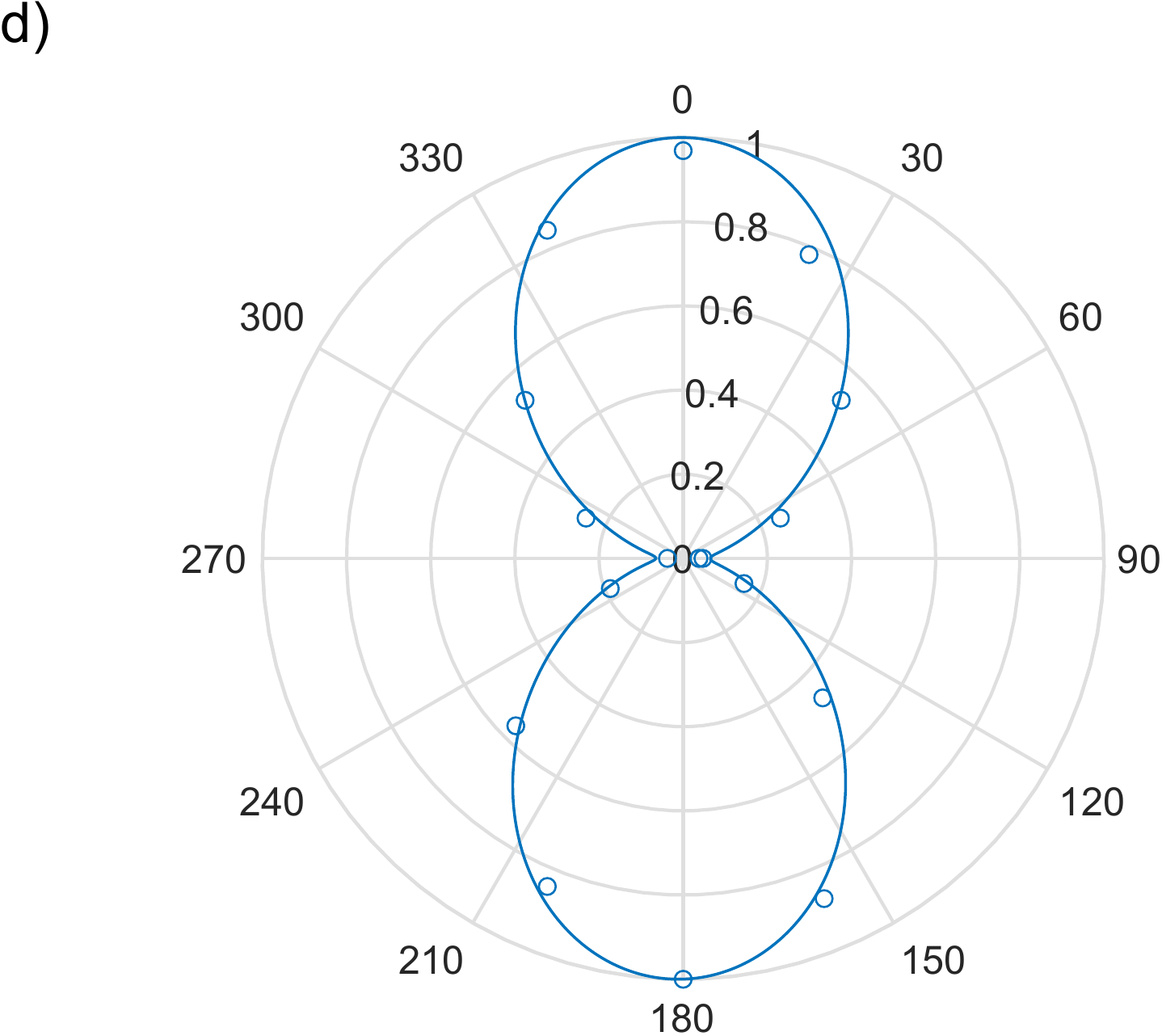}\label{sin2}
\caption{\label{satura} \textbf{Optical properties of single GeV$^-$ centres}. \textbf{\ref{satura}a} Saturation curves taken for  GeV1 and GeV2 centres show different brightness between emitters with background subtraction. \textbf{\ref{satura}b} Lifetime measurement under green pulsed laser for GeV1 and GeV2. \textbf{\ref{satura}c,\ref{satura}d} Emission polarization measurements for GeV1 and GeV2. All solid lines refer to the fitting model, see main text.}
\end{figure*}

\subsection{Fluorescence decay rate}
The rate coefficient analysis allows us to estimate both the excited and the metastable state lifetimes, expressed by $\tau_1= \frac{1}{k_{21}^0}$ and $\tau_2= \frac{1}{k_{31}^0}$ respectively. To validate the value of $\tau_1$ obtained by the model, we compare it to the experimental data obtained by a direct measurement of the lifetime using a pulsed laser. Figure \ref{satura}b shows a lifetime measurement under a 532 nm pulsed laser excitation at 10 MHz repetition rate. To validate the fluorescence decay of these emitters, the data were fitted using a single exponential model according to the equation $I=I_0+Ae^{-\frac{t}{\mid \tau \mid}}$. After normalisation, the model is simplified to $I=e^{-\frac{t}{\mid \tau \mid}}$. Using the fitted model, a total excited state lifetime of 10.11 ns and 20.5 ns for GeV1 and GeV2 respectively were found. These values are in agreement with those obtained using the three level model with intensity dependent de-shelving which are 9.25 ns and 19.58 ns for GeV1 and GeV2 respectively. The direct measurement of the lifetime reflects the total decay rate of the colour centre from its excited state. It includes the radiative and non radiative decay paths $\Gamma = \Gamma_{rad}+\gamma_{nr}$, where $\Gamma_{rad}$ represents the rate coefficients $k_{21}$. 
We point out that the shorter lifetime for GeV1 comes along with an increase of its brightness which will be shown in the following section. 


\subsection{Brightness of single GeV$^-$ centres}
 The brightness of the emitters is a very important figure of merit for a single photon source. Figure \ref{satura}a displays the saturation curves of the two single colour centres GeV1 and GeV2 taken after background correction. The solid fitted curves follow the equation  $I=I_{\infty}\frac{P}{P_{sat}+ P}$, where $I_{\infty}$ is the maximum emission intensity and $P_{sat}$ is the saturation power. 
Using the fitting curves, we measure a maximum emission intensity of 1.5 MHz and 0.2 MHz for GeV1 and GeV2 respectively with a saturation power of 0.56 and 1.5 mW. This indicates a 10-fold enhancement in the brightness of GeV$^-$ centres than that in bulk diamond reported elsewhere \cite{iwasaki2015germanium}. This enhancement is expected to be due to the very small size of our NDs. Indeed, the small size of the NDs results in an efficient suppression of total internal reflection, thus enabling us to collect more photons with the microscope objective \cite{beveratos2001nonclassical}. We point out here that these saturation curves are done on the ZPL line with a narrow band-pass filter with a FWHM of 14 nm around 600 nm. We thus expect that this value will increase if a wider band-pass is used. Moreover, our dry microscope objective has an NA=0.95 and will thus not collect as much as an oil-immersion objective would and therefore higher count rates are expected. Additional loss is due to the dichroic mirror used in our PL setup that suffers from a polarization-dependent loss of the emitted light.  
Therefore, it appears that the GeV$^-$ centres studied in this article have a higher quantum yield than SiV$^-$ in general, which has a similar saturation fluorescence but a lower excited state lifetime of less than 1 ns \cite{rogers2014multiple}. GeV1 showed a stable emission during a long time measurement, even at a very high excitation power. While most of the GeV$^-$ centres presented a stable emission, few others such as GeV2 have shown an unstable emission. This is due to the high coupling to the metastable state where the emitter goes in a `dark' or an `OFF' state. This is in agreement with the high value of $a$ obtained using our proposed model.

\subsection{Polarization of single GeV$^-$ centres}
In this paragraph, we investigate whether these single GeVs defects act as single dipole emitters. To this end, we check the polarization sensitivity by rotating a polarizer in the detection path. The variation of the emission intensity $I$ as a function of the angle $\theta$ is shown in figures \ref{satura}c and \ref{satura}d. We fit the data using equation $I=\alpha+\beta\sin^2(\theta + \phi)$, where $\alpha$, $\beta$ and $\phi$ are the fitting parameters. Inferred from the figures, we deduced that each centre possesses a preferred single linear polarization. The visibility or the degree of polarization calculated by $V=\frac{I_{max} -I_{min}}{I_{max} + I_{min}}$ is found to be around 92\%. This high polarisation visibility makes these GeV centres suitable for single photon source applications as well as for QKD applications which require photons with a well defined polarization state.

\section{CONCLUSION}
In this article, we show that single GeV$^-$ centres in very small NDs grown by HPHT display superior optical properties as compared to their bulk counterpart. Assuming that our emitters follow an extended three level system with intensity dependent de-shelving, we developed a model and optimised its parameters to fit the experimental data. The model was validated on several GeVs at different excitation powers. We have studied the single photon emission properties of the GeVs centres as well as their internal population dynamics represented by rate coefficients. We have also characterized their brightness, lifetime and polarization. The brightest GeV$^-$ studied had a maximum count rate of 1.6 MHz and acts as a single dipole emitter. As such, GeV$^-$ colour centres in such small NDs have demonstrated favorable properties for solid state single photon source applications. In future works, we will focus on studying the indistiguishability of the photons emitted from these GeV centres. In  addition, we aim to couple these single GeV centres in NDs to a photonic platform made of an ion exchange glass waveguide, towards the realization of a quantum integrated photonic circuit \cite{xu2020towards}.

\begin{acknowledgments}
 The authors are thankful to the European Union’s Horizon 2020 Research and Innovation Program under Marie Sklodowska-Curie Grant Agreement No. 765075 (LIMQUET). They also thank the platform  Nano'Mat (http://www.nanomat.eu) where experiments were carried out. V.D. and L.K. thank the Russian Foundation for Basic Research for financial support (Grant 18-03-00936).
\end{acknowledgments}

\section*{Author contributions}
M.N. carried out the experiments, performed the data acquisition, analyzed the data, and wrote the manuscript. D.A. developed the optimisation algorithm for the fitting models and wrote the Matlab codes. R.D. helped in the use of the instrumentation for the micro-PL experimental setup. V.D., L.K., and V.A. provided the GeV nanodiamond samples and performed the transmission electron microscopy (TEM) for the nanodiamond samples. C.C. conducted the research project, supervised the work technically, discussed the obtained results and edited the paper. All authors reviewed the manuscript.

\section*{References}
\bibliographystyle{ieeetr}
\bibliography{aipsamp}


\end{document}